\documentclass[12pt,a4paper,twoside]{article}

\usepackage[dvips]{graphics}
\usepackage{graphicx}
\usepackage{cite}

\def\be{\begin{equation}}
\def\ee{\end{equation}}
\def\bea{\begin{eqnarray}}
\def\eea{\end{eqnarray}}

\newcommand{\dd}       {\mathrm{d}}

\newcommand{\bbbar}     {\ensuremath{\mathrm{b\bar{b}}}}

\newcommand{\epem}              {\ensuremath{\mathrm{e^+e^-}}}

\newcommand{\nf}                {\ensuremath{n_{\mathrm{f}}}}
\newcommand{\as}                {\ensuremath{\alpha_{\mathrm{S}}}}
\newcommand{\asi}                {\ensuremath{\alpha_{\mathrm{S},i}}}
\newcommand{\asb}               {\ensuremath{\bar{\alpha}_{\mathrm{S}}}}
\newcommand{\asbsq}             {\ensuremath{\bar{\alpha}_{\mathrm{S}}^2}}

\newcommand{\asmz}              {\ensuremath{\alpha_\mathrm{S}(M_{\mathrm{Z^0}})}}

\newcommand{\oa}                {\ensuremath{\mathcal{O}(\alpha_\mathrm{S})}}
\newcommand{\oaa}               {\ensuremath{\mathcal{O}(\alpha_\mathrm{S}^2)}}

\newcommand{\evis}              {\ensuremath{E_{\mathrm{vis}}}}

\newcommand{\pmiss}     {\ensuremath{p_{\mathrm{miss}}}}
\newcommand{\pbal}              {\ensuremath{p_{\mathrm{bal}}}}

\newcommand{\momone}[1] {\mbox{\ensuremath{\langle#1\rangle}}}
\newcommand{\momn}[2] {\mbox{\ensuremath{\langle#1^{#2}\rangle}}}
\newcommand{\cp}                {\ensuremath{C}}
\newcommand{\bt}                {\ensuremath{B_\mathrm{T}}}
\newcommand{\bw}                {\ensuremath{B_\mathrm{W}}}

\newcommand{\mh}                {\ensuremath{M_\mathrm{H}}}

\newcommand{\thr}               {\ensuremath{1-T}}
\newcommand{\ytwothree}         {\ensuremath{y_{23}}}

\newcommand{\chisq}     {\ensuremath{\chi^2}}
\newcommand{\chisqd}    {\ensuremath{\chi^2/\mathrm{d.o.f.}}}
\newcommand{\xmu}               {\ensuremath{x_{\mu}}}

\newcommand{\ntrkl}             {\ensuremath{N_{\mathrm{long}}}}

\newcommand{\ycut}              {\ensuremath{y_{\mathrm{cut}}}}

\newcommand{\stat}              {\ensuremath{\mathrm{(stat.)}}}

\newcommand{\expt}               {\ensuremath{\mathrm{(exp.)}}}
\newcommand{\had}               {\ensuremath{\mathrm{(had.)}}}
\newcommand{\theo}              {\ensuremath{\mathrm{(theo.)}}}

\newcommand{\rs}                {\ensuremath{\sqrt{s}}}
\newcommand{\rsp}               {\ensuremath{\sqrt{s'}}}

\newcommand{\sigtot} {\ensuremath{\sigma_{\mathrm{tot}}}}

\newcommand{\signull}{\ensuremath{\sigma_{0}}}

\newcommand{\invpb}     {\ensuremath{\mathrm{pb}^{-1}}}

\newcommand{\bm}[1]     {\mbox{\boldmath\ensuremath{#1}}}

\newcommand{\py}                {PYTHIA}
\newcommand{\hw}                {HERWIG}
\newcommand{\ar}                {ARIADNE}

\newcommand{\cdet}    {\ensuremath{C^{\mathrm{detector}}}}
\newcommand{\chad}    {\ensuremath{C^{\mathrm{had}}}}

\newcommand{\result} {
\ensuremath{\asmz=0.1286\pm0.0007\stat\pm0.0011\expt\pm0.0022\had\pm0.0068\theo}}
\newcommand{\restot} {\ensuremath{\asmz=0.1286\pm0.0072~(\mathrm{total~error})}}

\begin{document}

\begin{titlepage}

\centerline{{\large Max-Planck-Institut f\"ur Physik}} \bigskip

\begin{flushright}
 JADE Note 147 \\
 MPP-2004-100 \\
 September 6, 2004
\end{flushright}

\bigskip\bigskip\bigskip

\begin{center}
\textbf{
\Large Study of moments of event shapes in \epem\ annihilation using JADE data
}
\end{center}

{\Large \par}

\bigskip

\begin{center}
  {\Large C. Pahl, S. Kluth, S. Bethke, P.A. Movilla Fernandez, J.
    Schieck, and the JADE Collaboration\footnote{See~\cite{naroska87}
      for the full list of authors}}
\end{center}

\par

\bigskip

\begin{abstract}
  Data from \epem\ annihilation into hadrons collected by the JADE
  experiment at centre-of-mass energies between 14~GeV and 44~GeV were
  used to study moments of event shape distributions.  The data were
  compared with Monte Carlo models and with predictions from QCD NLO
  order calculations.  The strong coupling constant measured from the
  moments is
\begin{center}
\result, \\
\restot \\
\end{center}
consistent with the world average. However, systematic deficiencies in
the QCD NLO order predictions are visible for some of the higher
moments.
\end{abstract}

\bigskip\bigskip\bigskip\bigskip\bigskip

{\large \par}

\bigskip\bigskip

\begin{center}
{\Large This note describes preliminary JADE results}
\end{center}

{\LARGE \par}

\vfill

\end{titlepage}

\section{Introduction}

The annihilation of an electron and a positron into hadrons allows
precise tests of Quantum Chromodynamics (QCD).  Commonly jet
production rates or distributions of event shape observables have been
studied.  Predictions by perturbative QCD combined with hadronization
corrections derived from models have been found to describe the data
at low and high energies well, see
e.g.~\cite{jadenewas,OPALPR299,movilla02b}.

In this analysis we used data collected by the JADE experiment in the
years 1979 to 1986 at the PETRA \epem\ collider at DESY at six
centre-of-mass energies \rs\ covering the range of 14--44~GeV.  We
measured the first five moments of event shape observables for the
first time in this \rs\ region in \epem\ annihilation and compared the
data to predictions by Monte Carlo Models and by perturbative QCD.
From the comparison of the data with the theory we extracted the
strong coupling constant \as.

The outline of the note is as follows. In section~2, we define the
observables used in the analysis and describe the
perturbative QCD predictions. In section~3 the analysis procedure is
explained in detail. Section~4 contains the discussion of the
systematic checks which were performed and the resulting systematic
errors. We collect our results in section~5 and summarize them in
section~6.

\section{Observables}
\label{theory}

The properties of hadronic events may be characterized by a set of
event shape observables.  These may be used to characterize the
distribution of particles and thus the topology of an event as
``pencil-like'', planar, spherical etc.  They can be computed either
using the measured charged particle tracks and calorimeter clusters,
or using the true hadrons or partons in simulated events.  The
following event shapes were considered here:
\begin{description}
\item[Thrust \bm{T}:]
  defined by the expression~\cite{thrust1,thrust2}
  \begin{displaymath}
  T= \max_{\vec{n}}\left(\frac{\sum_i|\vec{p}_i\cdot\vec{n}|}
                    {\sum_i|p_i|}\right)\;\;\;.
  \end{displaymath}
  The thrust axis $\vec{n}_T$ is the direction $\vec{n}$ which
  maximizes the expression in parentheses.  A plane through the origin
  and perpendicular to $\vec{n}_T$ divides the event into two
  hemispheres $H_1$ and $H_2$.
\item[\bm{C}-parameter:]
  The linearized momentum tensor $\Theta^{\alpha\beta}$ is defined by
  \[
    \Theta^{\alpha\beta}= \frac{\sum_i(p_i^{\alpha}p_i^{\beta})/|\vec{p}_i|}
                               {\sum_i|\vec{p}_i|}\;\;\;,
                           \;\;\;\alpha,\beta= 1,2,3\;\;\;.
  \]
  The three eigenvalues $\lambda_j$ of this tensor define
  \cp~\cite{ert} through
 \[
    \cp= 3(\lambda_1\lambda_2+\lambda_2\lambda_3+\lambda_3\lambda_1)\;\;.
  \]
\item[Heavy Jet Mass \bm{\mh}:] The hemisphere 
invariant masses are calculated using  the particles
  in the two hemispheres $H_1$ and $H_2$.   We define
  \mh~\cite{jetmass,clavelli81} as the heavier mass, divided by $\rs$.
\item[Jet Broadening observables \bm{\bt} and \bm{\bw}:] 
  These are defined by computing the quantity
  \[
    B_k= \left(\frac{\sum_{i\in H_k}|\vec{p}_i\times\vec{n}_T|}
                    {2\sum_i|\vec{p}_i|}\right)
  \]
  for each of the two event hemispheres, $H_k$,  defined above.
  The two observables~\cite{nllabtbw} are defined by
  \[
    \bt= B_1+B_2\;,\;\;\mathrm{and}\;\;\;\bw= \max(B_1,B_2)\;\;\;
  \]
  where \bt\ is the total and \bw\ is the wide jet broadening.
\item[Transition value between 2 and 3 jets {\boldmath \ytwothree}:]
 Jet algorithms are applied to cluster the large number of particles of
 an hadronic event into a small number of jets, reflecting the parton
 structure of the event. For this analysis we used the Durham
 scheme~\cite{durham}. Defining each particle initially to be a
 jet, a resolution variable $y_{ij}$ is calculated for each
 pair of jets $i$ and $j$: 
\be
  y_{ij}= \frac{2\mathrm{min}(E_i^2,E_j^2)}{E_{\mathrm{vis}}^2}
          (1-\cos\theta_{ij}),
\ee 
 where $E_{i}$ and $E_{j}$ are the energies, $\cos\theta_{ij}$ is the
 angle between the two jets and $E_{\mathrm{vis}}$ is the sum of the
 energies of all visible particles in the event (or the partons in a
 theoretical calculation).  If the smallest value of $y_{ij}$ is less
 than a predefined value \ycut, the pair is replaced by a jet with four
 momentum $p_{ij}^\mu = p_i^\mu + p_j^\mu$, and the clustering starts
 again with $p_{ij}^\mu$ instead of the momenta $p_i^\mu$ and
 $p_j^\mu$.  Clustering ends when the smallest value of $y_{ij}$ is
 larger than \ycut.  The remaining jets are then counted.
 The value of \ycut\ at which
 for an event the transition between a 2-jet and a 3-jet assignment
 occurs is called \ytwothree.
\end{description}
 
In the following discussion, whenever we wish to refer to a generic
event shape observable we use the symbol $y$.  In almost all cases,
larger values of $y$ indicate regions dominated by the radiation of
hard gluons and small values of $y$ indicate the region influenced by
multiple soft gluon radiation.  Note that thrust $T$ forms an
exception to this rule, as the value of $T$ reaches unity for events
consisting of two collimated back-to-back jets. We therefore use
$y=\thr$ instead.  For all of these event shapes, a perfectly
collimated (``pencil-like'') two-jet final state will have $y=0$.

The $n$th, $n=1,2,\ldots$, moment of the distribution of an event
shape observable $y$ is defined by
$$\momn{y}{n}=\int_0^{y_{max}} y^n \frac{1}{\sigma}
\frac{\dd\sigma}{\dd y} \dd y \;\;\;,$$
where $y_{max}$ is the
kinematically allowed upper limit of the observable.  The calculations
always involve a full integration over phase space, which implies that
comparison with data always probes all of the available phase space.
This is in contrast to QCD predictions for distributions; these are
commonly only compared with data, e.g.\ in order to measure \as, in
restricted regions, where the theory is able to describe the data
well, see e.g.~\cite{jadenewas}.  Comparisons of QCD predictions for
moments of event shape distributions with data are thus complementary
to tests of the theory using the differential distributions.

The formula for the \oaa\ QCD prediction of \momn{y}{n}\ is with
$\asb=\as/(2\pi)$
\begin{equation}
  \momn{y}{n} = {\cal A}_n \asb + {\cal B}_n \asbsq
\label{eq_qcdmom}
\end{equation}
involving the \oa\ coefficients ${\cal A}_n$ and \oaa\ coefficients
${\cal B}_n$.  The values of the coefficients ${\cal A}_n$ and ${\cal
  B}_n$ can be obtained by numerical integration of the QCD matrix
elements using the program EVENT2~\cite{event2}.  

The QCD prediction depends on the renormalization scale $\mu$, see
e.g.~\cite{ellis96}.  The renormalization scale factor is defined as
$\xmu=\mu/\rs$ implying that $\asb=\asb(\mu)$ in
equ.~(\ref{eq_qcdmom}).  A truncated fixed order QCD calculation such
as equ.~(\ref{eq_qcdmom}) will depend on \xmu.  The renormalization
scale dependence is implemented by the replacement ${\cal
  B}_n\rightarrow {\cal B}_n+\beta_0\ln\xmu{\cal A}_n$ where
$\beta_0=11-\frac{2}{3}\nf$ is the leading order $\beta$-function
coefficient of the renormalization group equation and $\nf=5$ is the
number of active quark flavours.

The data are normalized to the total hadronic cross section
$\sigtot=\signull(1+\as/\pi)$ in LO while the QCD calculations are
normalized to the born cross section \signull.  Thus a correction has
to be applied by making the replacement ${\cal B}_n\rightarrow {\cal
  B}_n-2{\cal A}_n$.

\section{Analysis Procedure}

\subsection{The JADE Detector}
\label{sec_detector}

A detailed description of the JADE detector can be found
in~\cite{naroska87}. This analysis relies mainly on the reconstruction
of charged particle trajectories and on the measurement of energy
deposited in the electromagnetic calorimeter.  Tracking of charged
particles was performed with the central detector, which was
positioned in a solenoidal magnet providing an axial magnetic field of
0.48 T. The central detector contained a large volume jet chamber.
Later a vertex chamber close to the interaction point and surrounding
$z$-chambers to measure the $z$-coordinate~\footnote{In the JADE
right-handed coordinate system the $+x$ axis pointed towards the
centre of the PETRA ring, the $y$ axis pointed upwards and the $z$
axis pointed in the direction of the electron beam. The polar angle
$\theta$ and the azimuthal angle $\phi$ were defined with respect to
$z$ and $x$, respectively, while $r$ was the distance from the
$z$-axis.}  were added. Most of the tracking information was obtained
from the jet chamber, which provided up to 48 measured space points
per track, and good tracking efficiency in the region $|\cos
\theta|<0.97$.  Electromagnetic energy was measured by the lead glass
calorimeter surrounding the magnet coil, separated into a barrel
($|\cos \theta|<0.839$) and two end-cap ($0.86<|\cos \theta|<0.97$)
sections.  The electromagnetic calorimeter consisted of 2520 lead
glass blocks with a depth of 12.5 radiation lengths in the barrel
(since 1983 increased to 15.7 in the middle 20\% of the barrel) and
192 lead glass blocks with 9.6 radiation lengths in the end-caps.

\subsection{Data Samples}

The data used in this analysis were collected by JADE between 1979 and
1986 and correspond to a total integrated luminosity of 195 \invpb.
The breakdown of the data samples, mean centre-of-mass energy, energy
range, data taking period, collected integrated luminosities and the size of the
data samples after selection of hadronic events are given in
table~\ref{lumi}.  The data samples were chosen following previous
analyses,
e.g.~\cite{naroska87,jadenewas,OPALPR299,movilla02b,pedrophd}.  The
data are available from two versions of the reconstruction software
from 9/87 and from 5/88.  We used both sets and considered differences
between the results as an experimental systematic uncertainty.

\begin{table}[htb!]
\begin{center}
\begin{tabular}{|r|r|r|r|r|r|} \hline
average       & energy       & year & luminosity  & selected & selected  \\
energy in GeV & range in GeV &      & (\invpb)    & events   & events  \\
 & & &                                            & 9/87     & 5/88 \\
\hline
14.0 & 13.0--15.0 & 1981       & 1.46 & 1722 & 1783 \\
\hline
22.0 & 21.0--23.0 & 1981       & 2.41 & 1383 & 1403 \\
\hline
34.6 & 33.8--36.0 & 1981--1982 & 61.7 & 14213 & 14313 \\
35.0 & 34.0--36.0 & 1986       & 92.3 & 20647 & 20876 \\
\hline
38.3 & 37.3--39.3 & 1985       & 8.28 & 1584 & 1585 \\
\hline
43.8 & 43.4--46.4 & 1984--1985 & 28.8 & 3896 & 4376 \\
\hline
\end{tabular}
\end{center}
\caption{
The average center-of-mass energy, energy range, year
of data taking and integrated luminosity for each data
sample, together with the numbers of selected data events using
the data sample version of 9/87 or 5/88.
}
\label{lumi}
\end{table}

\subsection{Monte Carlo Samples}

Samples of Monte Carlo simulated events were used to correct the data
for experimental acceptance and backgrounds. The process
$\epem\to\mathrm{hadrons}$ was simulated using \py~5.7~\cite{jetset3}.
Corresponding samples using \hw~5.9~\cite{herwig,herwig65} were used
for systematic checks.  The Monte Carlo samples generated at each
energy point were processed through a full simulation of the
JADE detector~\cite{jadesim1,jadesim2,jadesim3}, 
summarized in ~\cite{pedrophd}, and reconstructed in
essentially the same way as the data.
 
 In addition, for comparisons
with the corrected data, and when correcting for the effects of
fragmentation, large samples of Monte Carlo events without detector
simulation were employed, using the parton shower models \py~6.158,
\hw~6.2 and \ar~4.11~\cite{ariadne3}.  All models were adjusted to
LEP~1 data by the OPAL collaboration.

The \ar\ Monte Carlo generator is based on a color dipole mechanism
for the parton shower.  The \py\ and \hw\ Monte Carlo programs use the
leading logarithmic approximation (LLA) approach to model the emission
of gluons in the parton shower.  For the emission of the first hard
gluon, the differences between the LLA approach and a leading order
matrix element calculation are accounted for.  However, the emission
of gluons later on in the parton shower is based only on a LLA cascade
and is expected to differ from a complete matrix element calculation.
For this reason we do expect deviations in the description of the data
by the Monte Carlo models.  Recently new Monte Carlo generators have
been developed, which implement a more complete simulation of the hard
parton emission~\cite{kuhn00}.  However, the models are not yet tuned
to data taken at LEP and were therefore not considered in this
analysis.

\subsection{Selection of Events}

The selection of events for this analysis aims to identify hadronic
event candidates and to reject events with a large amount of energy
emitted by initial state radiation (ISR).  The selection of hadronic
events was based on cuts on event multiplicity (to remove leptonic
final states) and on visible energy and longitudinal momentum balance
(to remove radiative and two-photon events, $\epem \to \epem$
hadrons).  The cuts used are documented 
in~\cite{StdEvSel1,StdEvSel2,StdEvSel3} and summarized in 
a previous publication~\cite{jadenewas}.


Standard criteria were used to select good tracks and clusters of
energy deposits in the calorimeter for subsequent analysis.  
Charged particle tracks were required to have at least 20
hits in r-$\phi$ and at least 12 in r-$z$ in the jet chamber.  The
total momentum was required to be at least 50~MeV.  Furthermore, the
point of closest approach of the track to the collision axis was
required to be less than 5~cm from the nominal collision point in the
$x-y$ plane and less than 35~cm in the $z-$direction. 

In order to mitigate the effects of double counting of energy from
tracks and calorimeter clusters a standard algorithm was
adopted which associated charged particles
with calorimeter clusters, and subtracted the estimated contribution
of the charged particles from the cluster energy. 
Charged particle tracks were assumed to be pions while
the photon hypothesis was assigned to electromagnetic energy clusters.
Clusters in the electromagnetic calorimeter were required to 
have an energy exceeding
0.15~GeV after the subtraction of the expected energy deposit of any
associated tracks.
From all accepted tracks and clusters $i$ the visible energy
$\evis=\sum_i E_i$, momentum balance $\pbal=|\sum_i p_{z,i}|/\evis$ and
missing momentum $\pmiss=|\sum_i \vec{p}_i|$ were calculated.  To
charged particle tracks the pion mass was assigned while the mass
of clusters was assumed to be zero.

Hadronic event candidates were required to pass the following selection 
criteria:
\begin{itemize}
\item The total energy deposited in the electromagnetic calorimeter
  had to exceed 1.2~GeV (0.2~GeV) for $\rs<16$~GeV, 2.0~GeV (0.4~GeV)
  for $16<\rs<24$~GeV and 3.0~GeV (0.4~GeV) for $\rs>24$~GeV in the
  barrel (each endcap) of the detector.
\item The number of good charged particle tracks was required to be
  greater than three reducing $\tau^{+}\tau^{-}$ and two-photon
  backgrounds to a negligible level.
\item For events with exactly four tracks configurations with three
  tracks in one hemisphere and one track in the opposite hemisphere
  were rejected.
\item At least three tracks had to have more than 24 hits in $r-\phi$
  and a momentum larger than 500~MeV; these tracks are called long
  tracks.
\item The visible energy had to fulfill $\evis/\rs>0.5$.
\item The momentum balance had to fulfill $\pbal<0.4$.
\item The missing momentum had to fulfill $\pmiss/\rs<0.3$.
\item The z-coordinate of the reconstructed event vertex had to lie
  within 15~cm of the interaction point.
\item The polar angle of the thrust axis was
  required to satisfy $|\cos(\theta_{\mathrm T})|<0.8$ in order that
  the events be well contained in the detector acceptance.
\end{itemize}
The numbers of selected events for each \rs\ are shown in
table~\ref{lumi} for the two versions of the data.

\subsection{Corrections to the Data}
\label{detectorcorrection}

All selected tracks and the electromagnetic calorimeter clusters
remaining after correcting for double counting of energy as described
above were used in the evaluation of the event shape moments.  The
values of the moments after all selection cuts had been applied are said 
to be at the detector level.

In this analysis events from the process $\epem\rightarrow\bbbar$ were
considered as background, since especially at low \rs\ the large
mass of the b quarks and of the subsequently produced B hadrons will
influence values of the moments.  The QCD predictions were
calculated for massless quarks and thus we chose to correct our data
for the presence of \bbbar\ events.

For the determination of the moments we calculated the sums $\sum_i
y_{i,data}^n$ ($n=1,\ldots,5$) where $i$ denotes all selected events.
The expected contribution of \bbbar\ background events $\sum_i
y^n_{i,\bbbar}$ was subtracted from the observed $\sum_i
y^n_{i,data}$.  The effects of detector acceptance and resolution and
of residual ISR were then accounted for by a multiplicative correction
procedure.

Two sets of $\sum_i y^n_i$ were calculated from Monte Carlo simulated
signal events; the first, at the detector level, treated the Monte
Carlo events identically to the data, while the second, at the
hadron level, was computed using the true momenta of the stable
particles in the event\footnote{ All charged and neutral particles
with a lifetime larger than $3\times 10^{-10}$s were considered
stable.}, and was restricted to events where $\rsp$, the centre-of-mass
energy reduced due to ISR, statisfied $\rs-\rsp<0.15$~GeV.
The Monte Carlo
ratio of the hadron level to the detector level for each moment value
was used as a correction factor for the data.  The corrected sums were
then normalized by the expected total number of events yielding the
final values of \momn{y}{n}.  The expected total number of events was
calculated from the number of selected events in the data in the same
way as for the moments.

\begin{figure}[htb!]
\begin{center}
\includegraphics[width=1.0\textwidth]{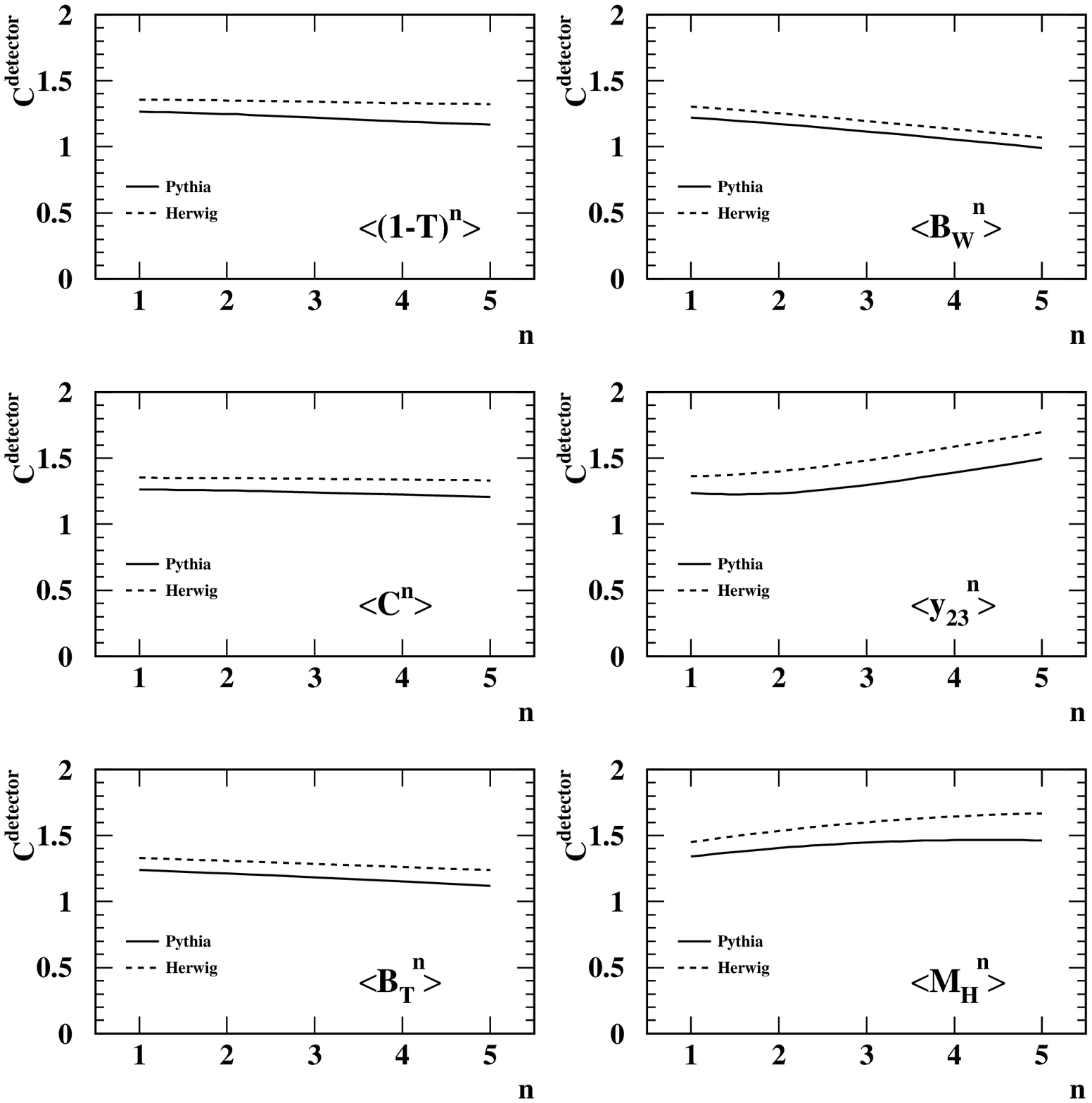}
\end{center}
\caption{The figures show the detector corrections 
at $\rs=14$~GeV as calculated using \py\ and \hw.}
\label{detcor}
\end{figure}

\begin{figure}[htb!]
\begin{center}
\includegraphics[width=1.0\textwidth]{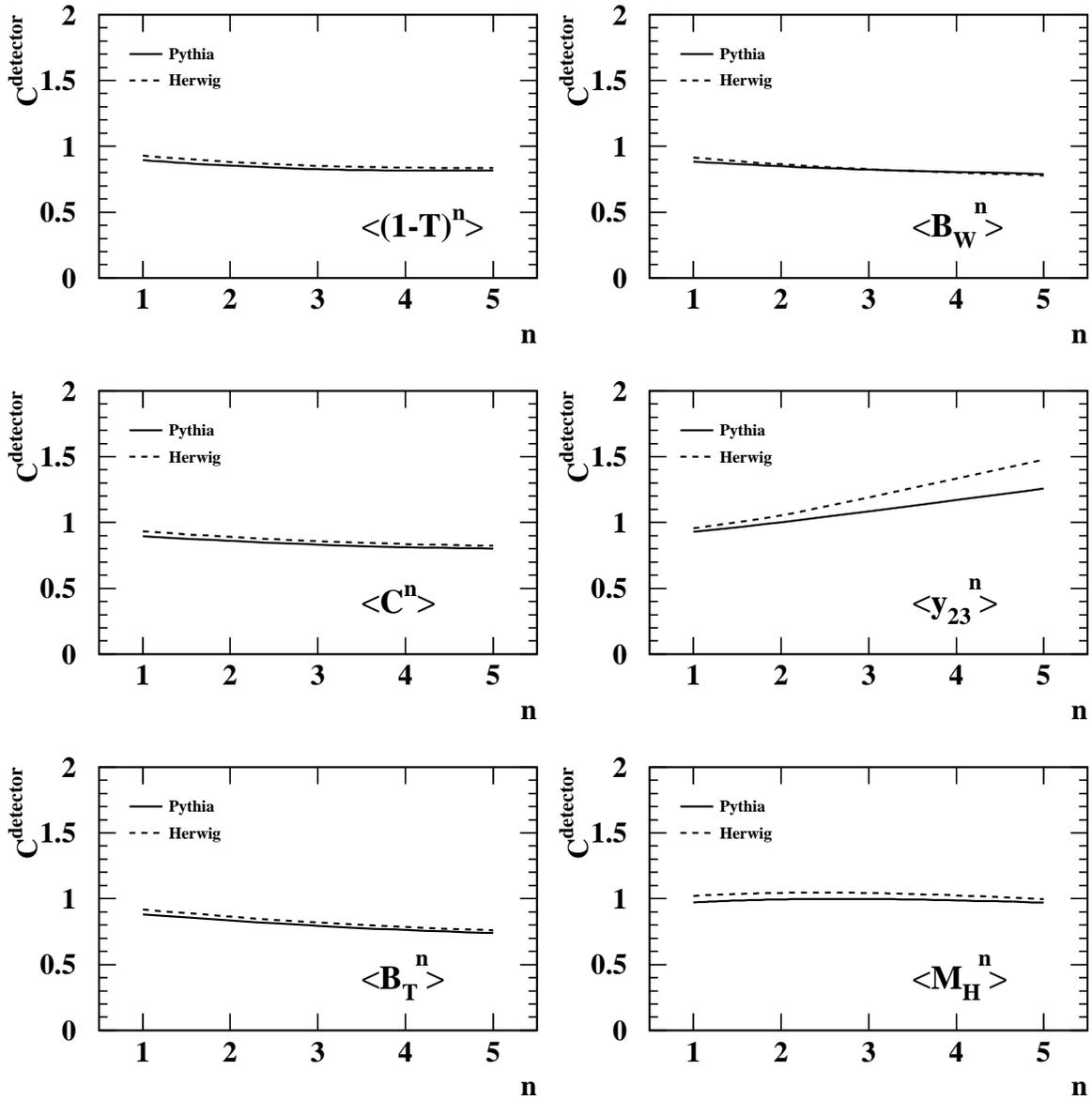}
\end{center}
\caption{Same as figure~\ref{detcor} for $\rs=35$~GeV }
\label{detcor2}
\end{figure}

The detector correction factors \cdet\ as determined using \py\ and
\hw\ are shown in figures~\ref{detcor} and~\ref{detcor2}.  We observe
some disagreement between the detector corrections calculated using
\py\ or \hw\ at low \rs\ while at larger \rs\ the correction factors
agree well for most observables.  The difference in detector corrections 
was evaluated as an experimental systematic uncertainty.

\section{Systematic Uncertainties}
\label{systematic}

Several sources of possible systematic uncertainties were studied.
All systematic uncertainties were taken as symmetric.  Contributions
to the experimental uncertainties were estimated by repeating the
analysis with varied cuts or procedures.  For each systematic
variation the value of \as\ was determined and then compared to the
result of the standard analysis (default value).  For each variation
the difference with respect to the default value was taken as a
systematic uncertainty.  In the cases of two-sided systematic
variations the larger deviation from the default value was taken as
the systematic uncertainty.
\begin{itemize}
\item In the standard analysis the data version from 9/87 was used.
  As a variation a different data set from 5/88 was used.
\item In the default method the tracks and clusters were
  associated and the estimated energy from the tracks was subtracted.
  As a variation all reconstructed tracks and all electromagnetic
  clusters were used.
\item The thrust axis was required to satisfy $|\cos(\theta_{\mathrm
    T})| < 0.7$.  With this more stringent cut events were restricted
  to the barrel region of the detector, which provides better
  measurements of tracks and clusters compared to the endcaps.
\item Instead of using \py\ for the correction of detector
  effects as described in section~\ref{detectorcorrection}, events
  generated with \hw\ were used.
\item The requirement on missing momentum was dropped or tightened to
  $\pmiss/\rs<0.25$.
\item The requirement on the momentum balance was dropped or tightened
  to $\pbal<0.3$.
\item The requirement on the number of long tracks was tightened to
  $\ntrkl\ge 4$.
\item The requirement on the visible energy was varied to $\evis/\rs>0.45$ and
  $\evis/\rs>0.55$.
\item The amount of subtracted \bbbar\ background was varied
  by $\pm$5\% in order to cover uncertainties in the estimation of
  the background fraction in the data.
\end{itemize}

All contributions listed above were added in quadrature and the result
quoted as the experimental systematic uncertainty.  The dominating
effects were the use of the different data version followed by
employing \hw\ to determine the detector corrections.  In the fits of
the QCD predictions to the data two further systematic uncertainties
were evaluated:

\begin{itemize}
\item The uncertainties associated with the hadronization correction
  (see section~\ref{fitprocedure}) were assessed by using \hw\ and
  \ar\ instead of \py. The larger change in \as\ resulting from these
  alternatives was taken to define the hadronization systematic
  uncertainty.
\item The theoretical uncertainty, associated with missing higher
  order terms in the theoretical prediction, was assessed by varying
  the renormalization scale factor \xmu.  The predictions of an
  all-orders QCD calculation would be independent of \xmu, but a
  finite order calculation such as that used here retains some
  dependence on \xmu.  The renormalization scale \xmu\ was set to 0.5
  and 2.  The larger deviation from the default value was taken as
  theoretical systematic uncertainty.
\end{itemize}

\section{Results}

\subsection{Values of Event Shape Moments}

\begin{figure}[htb!]
\begin{center}
\includegraphics[width=0.8\textwidth]{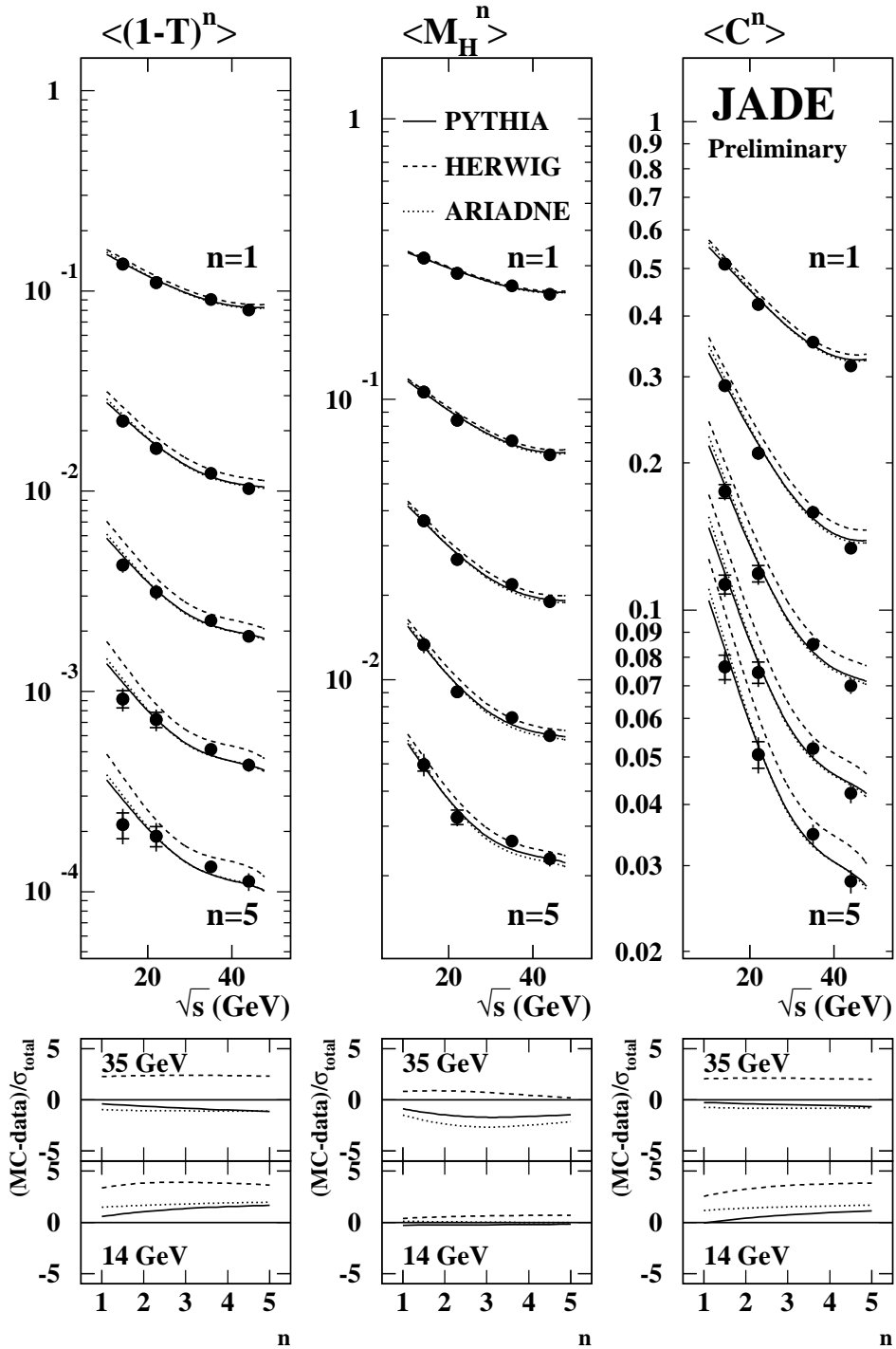} 
\end{center}
\caption{The figures show the first five moments of \thr, \mh\ and \cp\
  at hadron level compared with predictions based on \py, \hw\ and
  \ar\ Monte Carlo events.  The errors shown include all statistical
  and experimental uncertainties.  The lower panels show the
  differences between data and Monte Carlo at $\rs=14$~ and 35~GeV,
  divided by the quadratic sum of the statistical and experimental error.}
\label{hadron}
\end{figure}

\begin{figure}[htb!]
\begin{center}
\includegraphics[width=0.8\textwidth]{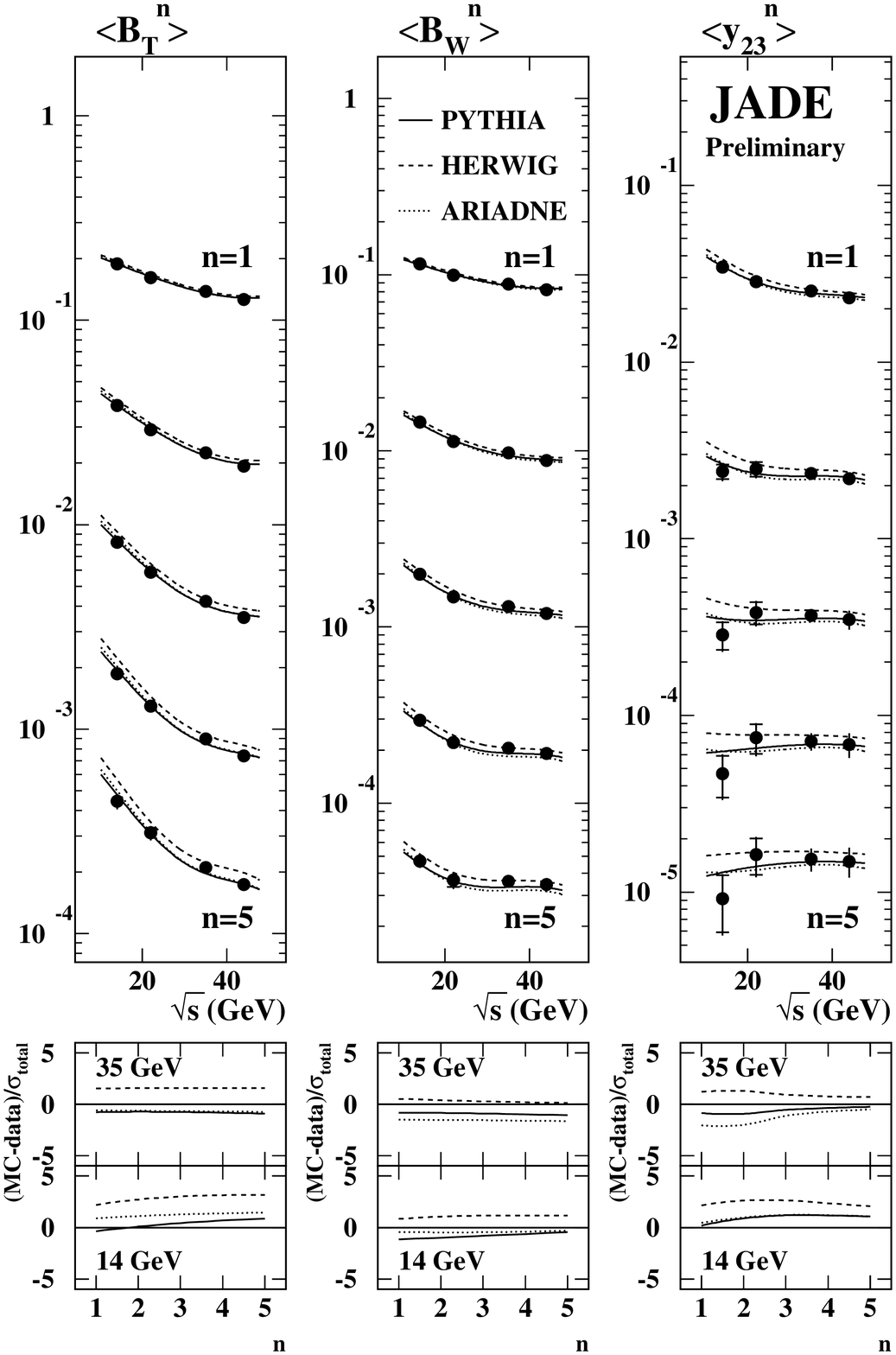}
\end{center}
\caption{ Same as figure~\ref{hadron} for \bt, \bw\ and \ytwothree. }
\label{hadron2}
\end{figure}

The first five event shape moments for the six observables after
subtraction of \bbbar\ background and correction for detector effects are
shown in figures~\ref{hadron} and~\ref{hadron2}.  Superimposed we show
the moments predicted by the \py, \hw\ and \ar\ Monte Carlo
models tuned by OPAL to LEP~1 data.  In order to make a more clear
comparison between data and models the lower plots show the
differences between data and each model divided by the combined
statistical and experimental error for $\rs=14$ and 35~GeV.  The three
models are seen to describe the data fairly well; \py\ and \ar\ are
found to agree better with the data than \hw.

\subsection{ Determination of \boldmath{\as} }
\label{fitprocedure}

Our measurement of the strong coupling constant \as\ is based on the
fits of QCD predictions to the corrected moment values, i.e. the data
shown in figures~\ref{hadron} and~\ref{hadron2}.  The theoretical
predictions using the \oaa\ calculation as described in
section~\ref{theory} provide values at the parton level.  In order to
confront the theory with the hadron level data, it is necessary to
correct for hadronization effects. The moments were calculated at
hadron and parton level using \py\ and, as a cross check, with the
\hw\ and \ar\ models.  The data points were then multiplied by the
ratio \chad\ of the parton and hadron level moment values in order to
correct for hadronization.

The models use cuts on quantities like e.g. the invariant mass between
partons in order to regulate divergencies in the predictions for the
parton shower evolution.  As a consequence in some events no parton
shower is simulated and the orginal quark-antiquark pair enters the
hadronization stage of the model directly.  This leads to a bias in
the calculation of moments at the parton level, since $y=0$ in this
case for all observables considered here.  In order to avoid this bias
we excluded in the simulation at the parton level events without at
least one radiated gluon.  At the hadron  and detector level all
events were used.

\begin{figure}[htb!]
\begin{center}
\includegraphics[width=1.0\textwidth]{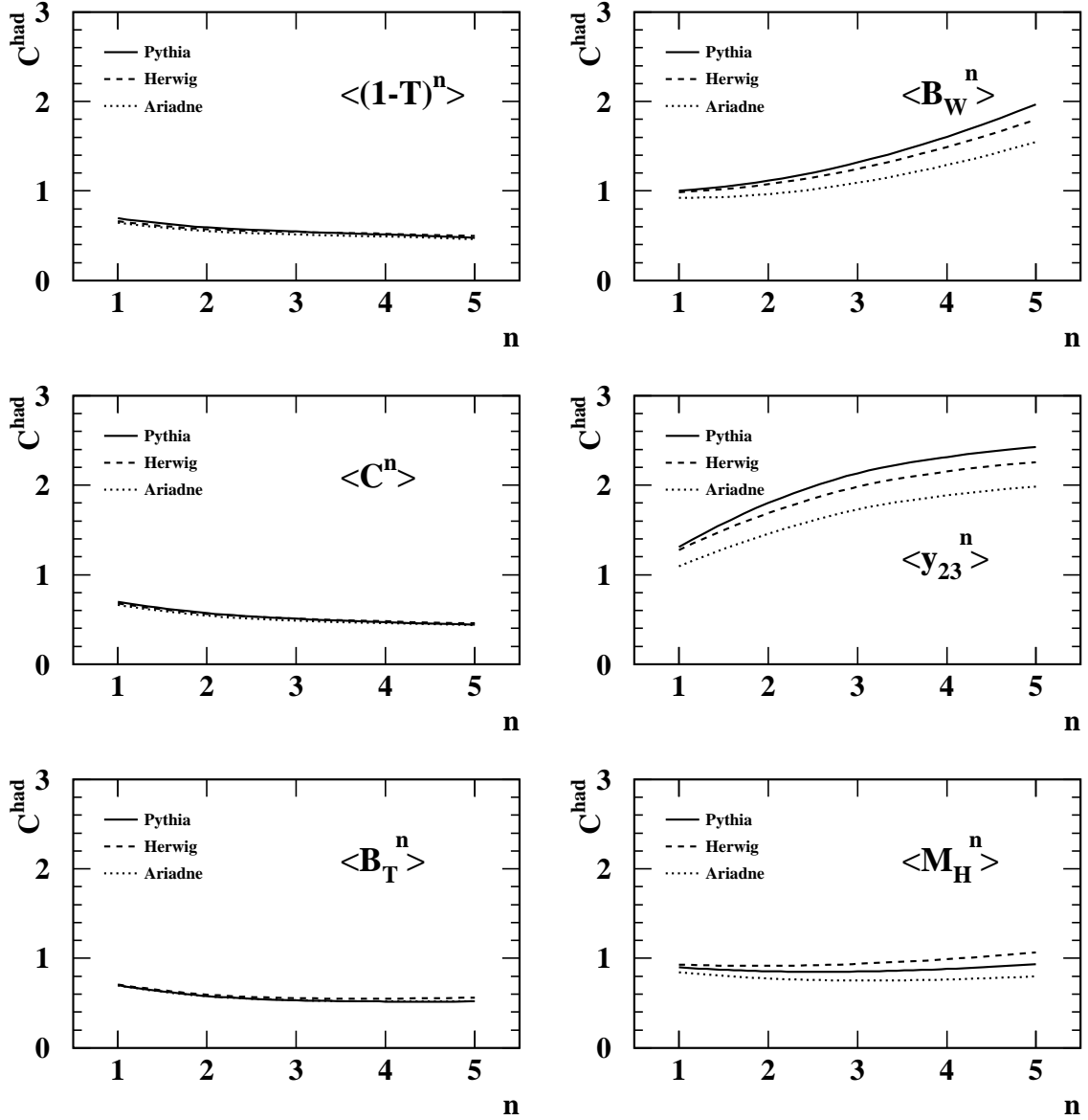}
\end{center}
\caption{The figures show the hadronization corrections at $\rs=14$~GeV
  as calculated using \py, \hw\ and \ar.}
\label{hadcor}
\end{figure}

\begin{figure}[htb!]
\begin{center}
\includegraphics[width=1.0\textwidth]{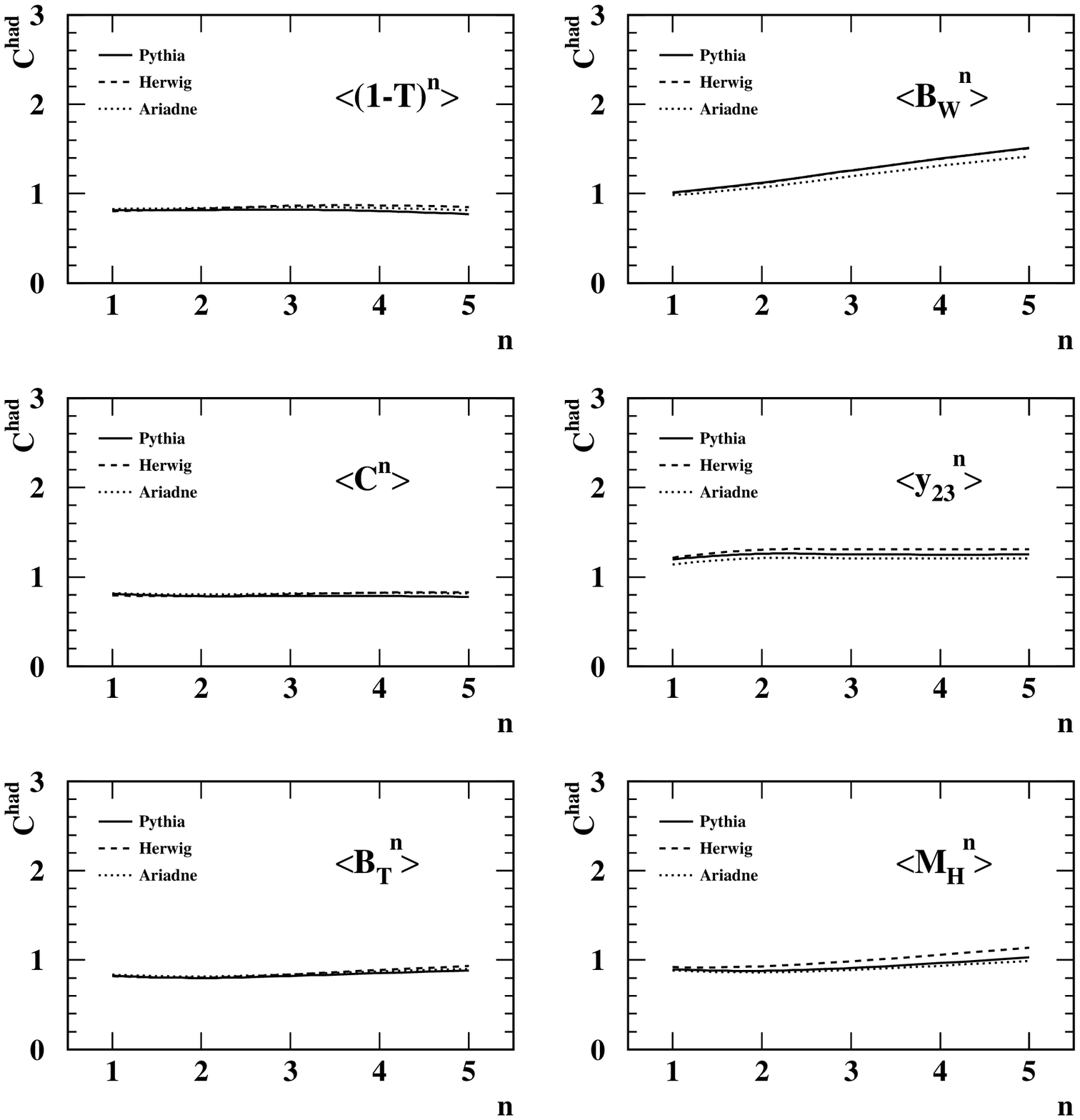}
\end{center}
\caption{Same as fig.~\ref{hadcor} for $\rs=35$~GeV.}
\label{hadcor2}
\end{figure}

The hadronization correction factors \chad\ as obtained from the three
models are shown in figures~\ref{hadcor} and~\ref{hadcor2}.  We find
that the hadronization corrections reach values of down to about 0.5 at
low \rs.  For larger \rs\ the hadronization corrections decrease as
expected.  We also observe that the models don't agree well for
moments based on \bw, \ytwothree\ and \mh\ at low \rs.  The
differences between the models were studied as a systematic
uncertainty in the fits.

A $\chi^{2}$-value for each moment \momn{y}{n} was calculated using the
following formula:
\be
 \chi^{2} = \sum_i (\momn{y}{n}_i-\momn{y}{n}_{\mathrm{theo},i})^{2}/\sigma_{i}^{2}
\ee 
where $i$ denotes the energy points.  The \chisq\ value was
minimized with respect to \asmz\ for each moment $n$ separately.  The
running of \as\ was implemented in the fit in two-loop precision using
the formula shown in~\cite{pdg04}.  The scale parameter \xmu, as
discussed in section~\ref{theory}, was set to 1.

\begin{figure}[htb!]
\begin{center}
\includegraphics[width=0.8\textwidth]{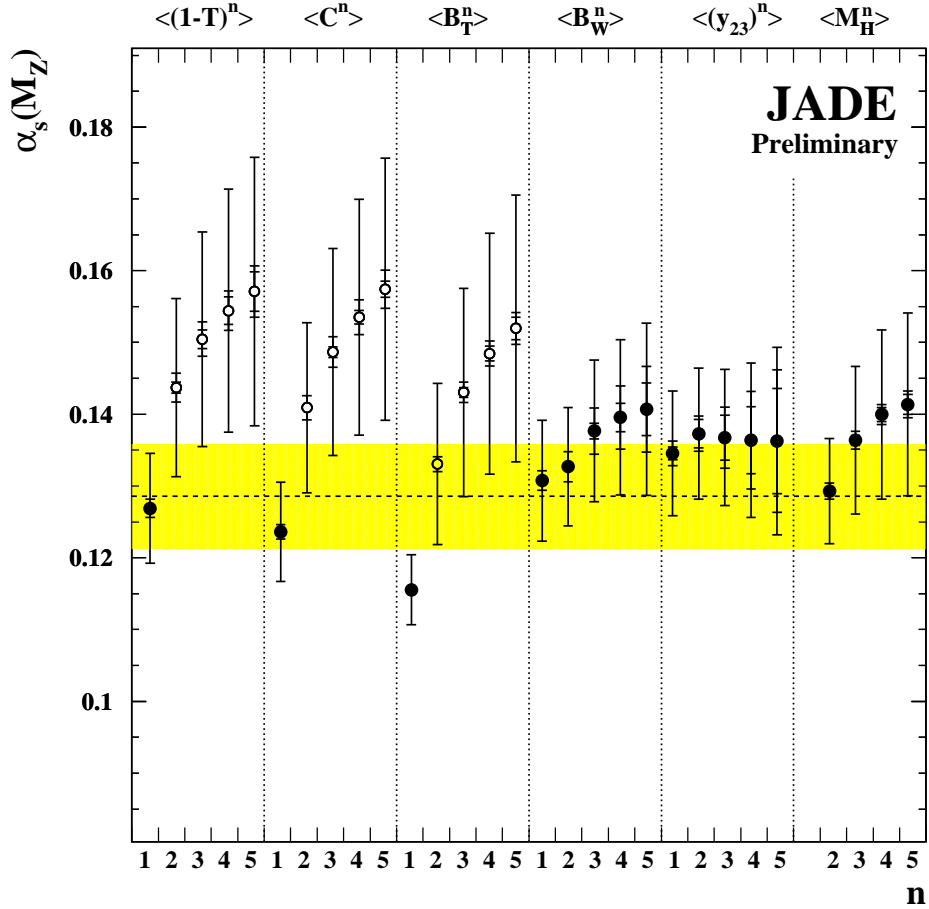} 
\end{center}
\caption{ Measurements of \asmz\  using fits to moments of six event
  shape observables.  The inner error bars represent statistical
  errors, the middle error bars include experimental errors and the
  outer error bars show the total errors.  The dotted line indicates
  the weighted average described in the text; only the measurements
  indicated by solid symbols were used for this purpose. }
\label{fit_plot}
\end{figure}

\begin{figure}[htb!]
\begin{center}
\includegraphics[width=0.8\textwidth]{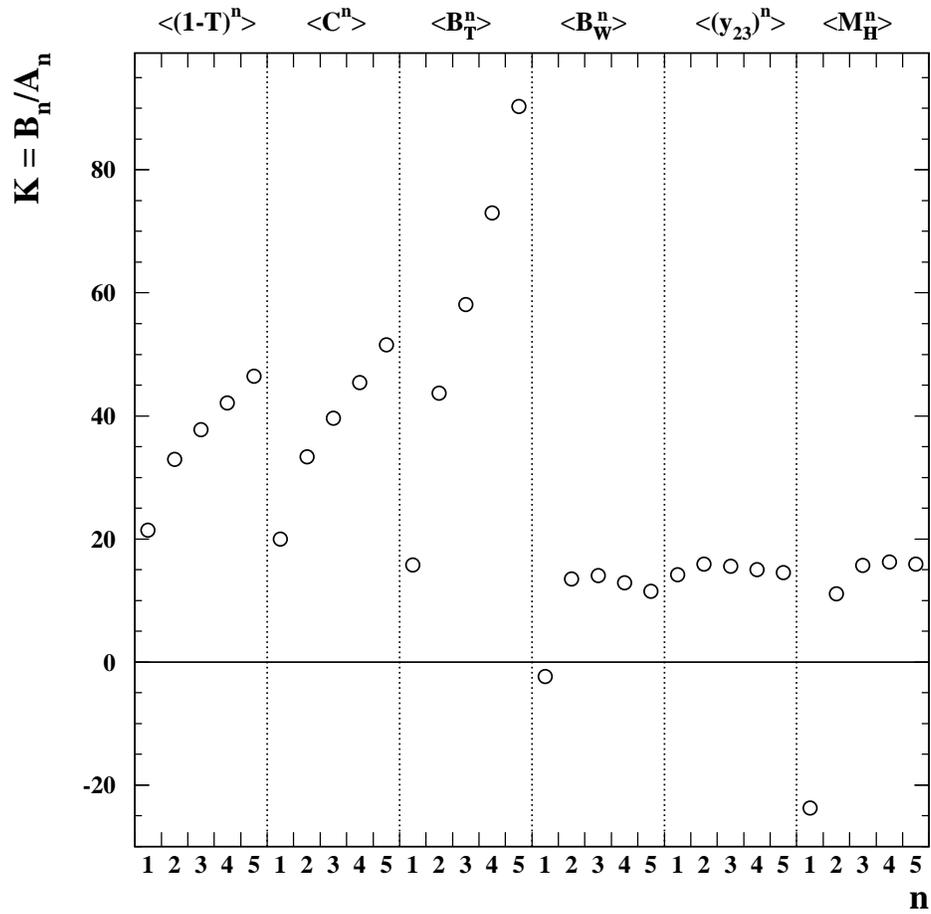} 
\end{center}
\caption{ The ratio $K={\cal B}_n/{\cal A}_n$ of NLO and LO coefficients
for the six observables.}
\label{baplot}
\end{figure}

The fit results are shown in figure~\ref{fit_plot}.  The fit to the
first moment \momone{\mh} did not converge and therefore no result is
shown.  We observe values of \chisqd\ of $\mathcal{O}(1)$; the fitted
QCD predictions including the running of \as\ are thus consistent with
our data.  However, we find that for \momn{(1-T)}{n}, \momn{\cp}{n}
and \momn{\bt}{n} the fitted values of \asmz\ increase steeply with
the order $n$ of the moment used.  This effect is not as pronounced
for \momn{\bw}{n}, \momn{(\ytwothree)}{n} and \momn{\mh}{n},
$n=2\ldots5$.  In order to investigate the origin of this behaviour we
show in figure~\ref{baplot} the ratio $K={\cal B}_n/{\cal A}_n$ of NLO
and LO coefficients for the six observables used in our fits.  There
is a clear correlation between the steeply increasing values of \asmz\ 
with moment $n$ and increasing values of $K$ with $n$ for
\momn{(1-T)}{n}, \momn{\cp}{n} and \momn{\bt}{n}.  The other
observables \momn{\bw}{n}, \momn{(\ytwothree)}{n} and \momn{\mh}{n},
$n=2\ldots5$, have fairly constant values of $K$ and correspondingly
stable results for \asmz.  We also note that \momone{\mh} has a large
and negative value of $K$ which is the cause that the fit did not
converge.

It is also of interest to combine the measurements of \asmz\ from the
various fits in order to determine a single value.  This problem has
been subject of extensive study by the LEP QCD working
group~\cite{LEPQCDWG}, and we adopt their procedure here.

In brief the method is as follows.  The measurements of \asmz\ were
combined in a weighted mean, to minimize the $\chi^{2}$ between the
combined value and the measurements.  If the measured values of \asmz\ 
are denoted \asi, with covariance matrix $V^{\prime}$, the combined
value, \asmz, is given by
\be 
  \asmz=\sum w_{i} \asi \;\;\;\; \mathrm{where}\;\;\;\;
  w_{i}=\frac{\sum_{j}(V^{\prime~-1})_{ij}}{\sum_{j,k}(V^{\prime~-1})_{jk}},
\ee
where $i$ and $j$ denote the individual results.  The difficulty
resides in making a reliable estimate of $V^{\prime}$ in the presence
of dominant and highly correlated systematic errors.  Small
uncertainties in the estimation of these correlations can cause
undesirable features such as negative weights.  For this reason only
statistical correlations and experimental systematic errors assumed to
be partially correlated between measurements were taken to contribute
to the off-diagonal elements of the covariance matrix:
$V^{\prime}_{ij}=\min(\sigma^2_{exp,i},\sigma^2_{exp,j})$.  All error
contributions (statistical, experimental, hadronization and scale
uncertainty) were taken to contribute to the diagonal elements.  The
hadronization and scale uncertainties were computed by combining the
\asmz\ values obtained with the alternative hadronization models, and
from setting $\xmu=0.5$ or $\xmu=2.0$, using the weights derived
from the covariance matrix $V^{\prime}$.

We considered only those results for which the NLO term in
equ.~(\ref{eq_qcdmom}) is less than half the LO term (i.e.\ 
$|K\as/2\pi|<0.5$ or $|K|<25$), namely
\momone{\thr}, \momone{\cp}, \momone{\bt}, \momn{\bw}{n} and
\momn{(\ytwothree)}{n}, $n=1,\ldots,5$ and \momn{\mh}{n},
$n=2,\ldots,5$; i.e.\ results from 17 observables in total.  The
purpose of this requirement was to select observables with an
apparently converging perturbative prediction.  The $K$ values are
shown in figure~\ref{baplot}.  The statistical correlations between the
17 results were determined using Monte Carlo simulation at the
hadron level.

The result of the combination is
\begin{displaymath}
 \result\;,
\end{displaymath}
above but still consistent with the world average value of
$\asmz=0.1182\pm0.0027$ ~\cite{bethke04}.  It has been observed previously
in comparisons of distributions of event shape observables with NLO
QCD predictions with $\xmu=1$ that fitted values of \asmz\ tend to
be large, see e.g.~\cite{OPALPR075}.

Combining only the fit results from \momone{\thr}, \momone{\cp}, 
\momone{\bt}, \momone{\bw}, \momone{\ytwothree} and \momn{\mh}{2} yields a 
value of
\begin{displaymath}
  \asmz=0.1239\pm0.0004\stat\pm0.0008\expt\pm0.0009\had\pm0.0059\theo \;.
\end{displaymath}
The slightly smaller error of \as\ reflects the fact that the lower
order moments are less sensitive to the multijet region of the 
event shape distributions. This leads to a smaller systematic uncertainty.

\section{Summary}

In this note we present preliminary measurements of the strong
coupling constant from moments of event shape distributions at centre-of-mass
energies between 14 and 44~GeV using data of the JADE experiment.  The
predictions of the \py, \hw\ and \ar\ Monte Carlo models tuned by OPAL
to LEP~1 data are found to be in reasonable agreement with the measured
moments.

From a fit of \oaa\ predictions to selected event shape moments
corrected for experimental and hadronization effects we have
determined the strong coupling constant to be \restot.
The higher moments, in particular of the Thrust, C-Parameter and
\bt\ event shape, lead to systematically enlarged value of \as.

\clearpage

\section*{ Figures }

\end{document}